\documentclass[preprintnumbers,twocolumn,showpacs,amsmath,amssymb]{revtex4}

\usepackage{graphicx}
\usepackage{dcolumn}
\usepackage{bm}

\begin{document}
\title{The twenty-four near-instabilities of Caspar-Klug viruses}
\author{Fran\c{c}ois Englert$^a$, Kasper 
Peeters$^{b}$, Anne
Taormina$^c$}
\affiliation{$^a$Service de Physique Th\'eorique and The International Solvay Institutes, Universit\'e Libre de Bruxelles,   Belgium\\
$^b$ Institute for Theoretical Physics, Utrecht University,
The Netherlands\\
$^c$ Department of Mathematical
Sciences, Durham University, United Kingdom}

\preprint{ULB-TH/08-10, ITP-UU-08/22, SPIN-08/20, DCPT-08/29}
\date{April 27th, 2008}
 
\begin{abstract}
Group theoretical arguments combined with normal mode analysis
techniques are applied to a coarse-grained approximation of
icosahedral viral capsids which incorporates areas of variable
flexibility. This highlights a remarkable structure of the
low-frequency spectrum in this approximation, namely the existence of
a plateau of 24 near zero-modes with universal group theory content.
\end{abstract}

\maketitle

\section{Introduction and summary}

Proteins in a thermal bath exhibit a wide spectrum of dynamical
behaviours which can be probed, for instance, by inelastic neutron
diffusion (see~\cite{Smith:2000} for a review).  In particular, they
undergo slow, large-amplitude motions which are now widely believed to
be instrumental to their function. These ideas were first tested
in~\cite{Karplus:1976, Noguti:1982a, Brooks:1983a, Go:1983,
  Levitt:1983, Harrison:1984} where normal mode analysis (NMA) was
used to argue that only a few low-frequency normal modes of vibration
are sufficient to describe, with great accuracy, the conformational
changes in a variety of proteins. The method has limitations, as it
assumes the existence of a single potential well whose minimum is a
given stable conformation of the protein under study, and therefore
overlooks the possibility of neighbouring multi-minima of energy
reported to exist in~\cite{Karplus:1987,Frauenfelder:1988, Hong:1990}.
Yet, this technique yields dynamical data which are consistent with
experimental results on proteins, as observed on case-by-case studies
in~\cite{Brooks:1985, Gibrat:1990, Marques:1995, Mouawad:1996}, and is
also confirmed by a recent statistical
analysis~\cite{Alexandrov:2005}. Although biologically significant
low-frequency motions are typically {\em not} vibrational due to the
damping influence of the protein environment, NMA captures the
tendency of the biomolecule to change in a few particular directions
corresponding to low-frequency normal modes, and thus remains a useful
tool when time-dependent methods like molecular dynamics are
prohibitive.

The success of NMA in studying protein dynamics has prompted its use
in the context of large macro-biomolecular assemblies. The main
motivation so far has been to pin down whether several experimentally
observed conformations of viral particles could be inferred from one
another by arguing that conformational changes occur in directions
which maximally overlap with those of a few low-frequency putative
normal modes of vibration of the capsid~\cite{
  Simonson:1992,Tama:2002a,Vlijmen:2005a, Rader:2005a}.  The biggest
challenge remains the choice, within the NMA framework, of a potential
which optimally captures the physics of capsid vibrations whilst
taking into account a reduced number of degrees of freedom to enable
practical calculations.  Many NMA applied to viruses implement
variations of the simple Elastic Network Model proposed a decade
ago~\cite{Tirion:1996a}, in which the atoms are taken as point masses
connected by springs modelling interatomic forces, provided the
distance between them is smaller than a given cutoff
parameter. Simplified versions include the restriction to
$C^{\alpha}$-atoms only, the approximation in which each residue is
considered as a point mass, or where even larger domains within the
constituent coat proteins are treated as rigid
blocks~\cite{Tama:2002a}. In an effort to optimise the NMA techniques
when applied to particles with high symmetry, group theoretical
considerations have also been
exploited~\cite{Simonson:1992,Vlijmen:2005a}.  When combined with an
Elastic Network approach, it allows for more extensive normal modes
calculations of viral capsids~\cite{Vlijmen:2005a} and compares well
with results obtained using an Elastic Network-RTB
setup~\cite{Tama:2002a}.

The elastic potential in all analyses above has two major drawbacks:
it does not discriminate between strong and weak bonds as it depends
on a single spring constant, and it uses the rather crude technique of
increasing the distance cutoff to resolve capsid
instabilities. Consequently, the frequency spectra have much less
structure than one would expect in reality, and in particular fail to
reproduce areas of rigidity and flexibility of the capsid
satisfactorily. This problem has been addressed in~\cite{Kim:2006a}
where a bond-cutoff method is implemented, together with different
spring constants for the various types of chemical interactions. 
The proposed model
reproduces conformational changes better than the conventional
distance-cutoff simulations.

In this paper, we propose a reductionist approach to the study of
icosahedral viral capsid vibrations within the harmonic approximation.
We consider Caspar-Klug viral capsids, which are classified
according to a triangulation number $T$~\cite{Caspar:1962a}. A $T=n$
capsid \footnote{$n=h^2+hk+k^2$ for $h$ and $k$ integer and
  relatively prime.} exhibits $60n$ coat proteins, organised in
clusters of twelve pentamers located at the vertices of an
icosahedron, and $10(n-1)$ hexamers at global 3-fold and/or local
6-fold symmetry axes of the icosahedral capsid. In our
coarse-graining, these coat proteins are approximated by point masses
located at their centres of mass, calculated using the Protein Data
Base (PDB).  We set up an elastic network whose representatives are
these point masses, with all masses normalised to 1. The network
topology is determined by data from the VIPER
website~\cite{Shepherd:2006a}: two point masses are connected by a
spring whenever VIPER provides a value for the association energy of
the two corresponding proteins. The spring constants of our model are
of the form $\kappa_{mn}=\rho_{mn} \kappa$, with $\rho_{mn}$ the ratio
of the association energy of protein pair $(m,n)$ to the largest
association energy listed in VIPER, and $\kappa$ is a free parameter
which reflects the lack of confidence in the {\em absolute} values of
association energies published in VIPER. The potential reads,
\begin{equation} 
V= \displaystyle{\sum_{\substack{m <n\\m,n =1}}^N\,\frac{1}{2} \kappa_{mn}\,(|\vec{r}_m-\vec{r}_n|-|\vec{r}^{\,0}_m-\vec{r}^{\,0}_n|)^2}\, ,
\end{equation}
where the vector $\vec{r}^{\,0}_m$ refers to the equilibrium position
of protein $m$, and the vector $\vec{r}_m$ to its position after
elastic displacement, all vectors originating from the centre of the
capsid. The resulting force (stiffness) matrix thus depends on the
parameter $\kappa$, and possesses more than the six zero eigenvalues
expected from the rotations and translations of the whole capsid
whenever the number of constraints imposed by the association energies
is smaller than $180n-6$, where $180n$ is the total number of degrees
of freedom for a $T=n$ capsid.  

Our main result, which will be substantiated by group theory arguments
below, is the existence, in the low-frequency spectrum of all our
coarse-grained stable Caspar-Klug capsids, of 24 near-zero normal
modes of vibration which always fall in the same set of non-singlet
irreducible representations of the icosahedral group
(see~\eqref{zeromodes}). The first singlet representation, which is
associated with a fully symmetric mode, always appears higher up in
the spectrum, in accordance with the expectation that such a motion
requires more energy to develop.  The presence of 24 near-zero modes
in the spectrum of viral capsids is deeply rooted in the fact that the
latter exhibit icosahedral symmetry.  Although our model is too crude
to account for a realistic flexibility of the viral capsid, and
therefore too primitive to get quantitative information on the virus
function, it can serve as a first step in an analysis which should
bridge the gap with existing molecular dynamics simulations. It
discriminates between strong and weak inter-protein bonds and
therefore captures effects of varying flexibility across the
capsid. Moreover, it is simple enough to highlight universal aspects
of the low-frequency spectrum of normal modes of vibration, inherited
from the underlying icosahedral symmetry, and to pin down the
influence of the elastic network design on the instability of the
capsid. 

Our results should thus be viewed as a remarkable property of the
lowest order approximation to virus capsid vibrations, which may
potentially be relevant once subsequent orders, with additional
interactions and additional degrees of freedom, are taken into
account. A meaningful comparison with existing virus NMA computations
in the literature is difficult at this stage, partly because of
reasons highlighted in~\cite{Kim:2006a}
regarding the use of Tirion's potential and the distance-cutoff
method.  However, it is noteworthy that the first non-zero mode (a
singlet of ${\cal I}$) in the spectrum of HK97, as calculated
in~\cite{Rader:2005a} using an all-atom simulation with a
cutoff, appears at mode 31 and is therefore compatible with our
qualitative arguments on the number of low-frequency
modes~\footnote{For the simpler icosahedrically symmetric~$C_{60}$, our
  approximation would involve 60 atoms connected by 90 bonds.
  Our prescription for counting the number of non-trivial zero modes 
  of this system suggests a plateau of 84~non-trivial low-frequency
  modes, rather than 24. However, it is unlikely that this result can
  be used to argue that, in the presence of a realistic potential, one
  should observe a gap in the spectrum. The 84 zero modes now span
  half of the total number of states, which is a much larger fraction
  than the zero modes of a $T=n$ capsid ($\frac{1}{6n}$), especially
  for $n > 1$. As each subsequent zero mode will be lifted by a small
  amount, perturbation theory is likely to become invalid as one
  approaches the end of the plateau.}.

\section{Low-frequency plateau}

The remainder of this paper provides the group theoretical arguments
leading to the 24 near zero-mode plateau in our coarse-grained
capsids.  We first approximate a Caspar-Klug $T=n$ capsid by merging
the $3n$ proteins per icosahedral face into a single point mass whose
equilibrium position is at the centre of the face. These point masses
are connected by springs (masses and spring constants normalised to~1)
between nearest neighbours, forming a dodecahedral cage with 30 edges,
dual to the icosahedron. Such a structure is unstable, since the
number of genuine degrees of freedom is $3 \times 20 -6 = 54$, while
there are only 30 constraints. Accordingly, the cage develops 24
non-trivial zero modes. To analyse these instabilities, we compare all
possible motions of this dodecahedral cage consistent with icosahedral
symmetry, with the dodecahedral motions induced by all possible
motions of the 12 vertices of an icosahedron. By induced, we mean that
the dual dodecahedron moves in such a way that its vertices are
located at the centres of the (deformed) icosahedral faces at all
times.

\begin{figure}[t]
\begin{center}
(a)\includegraphics[width=6.2cm,keepaspectratio]{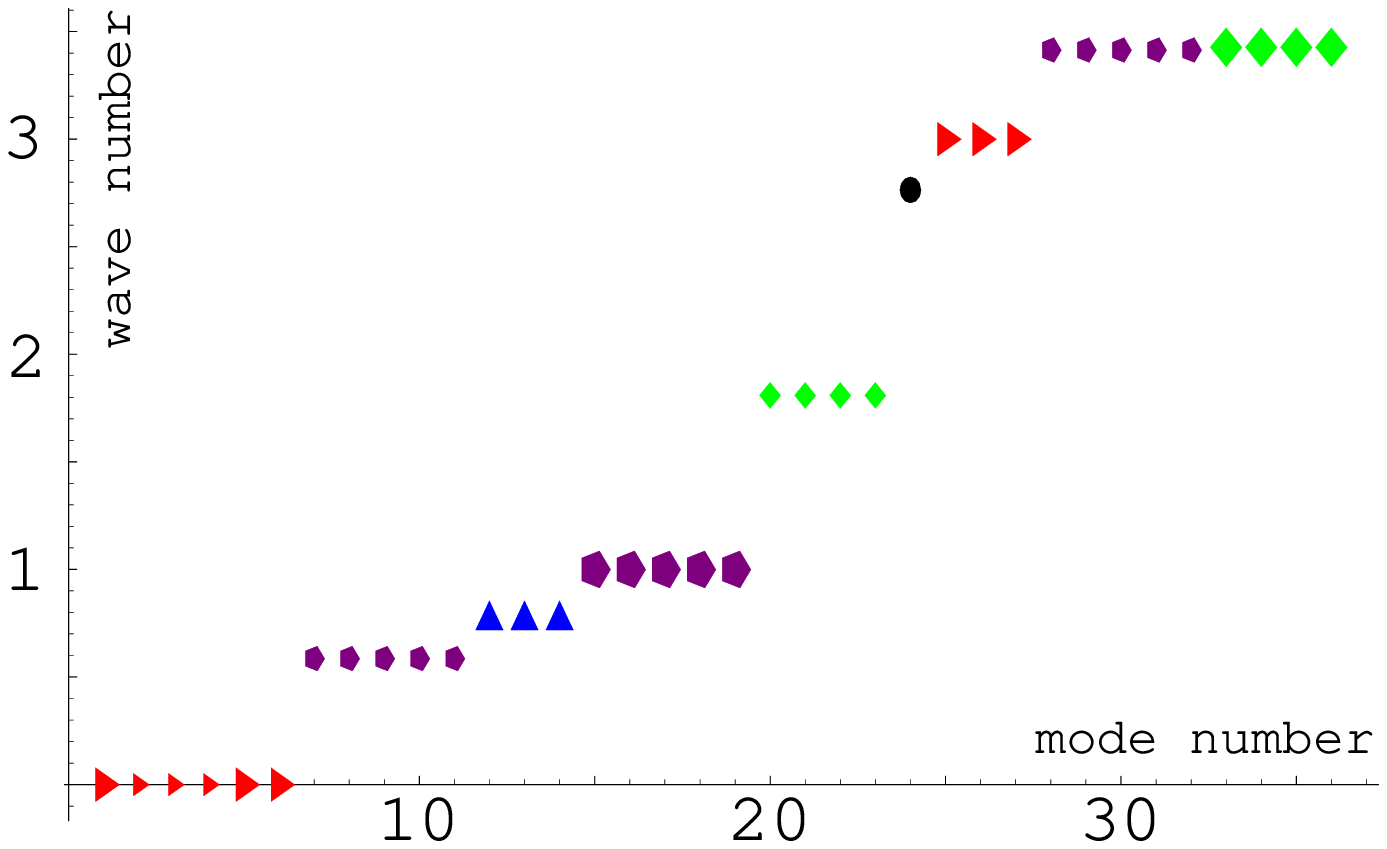}\\
\vspace{.5 cm}
(b)\includegraphics[width=6.2cm,keepaspectratio]{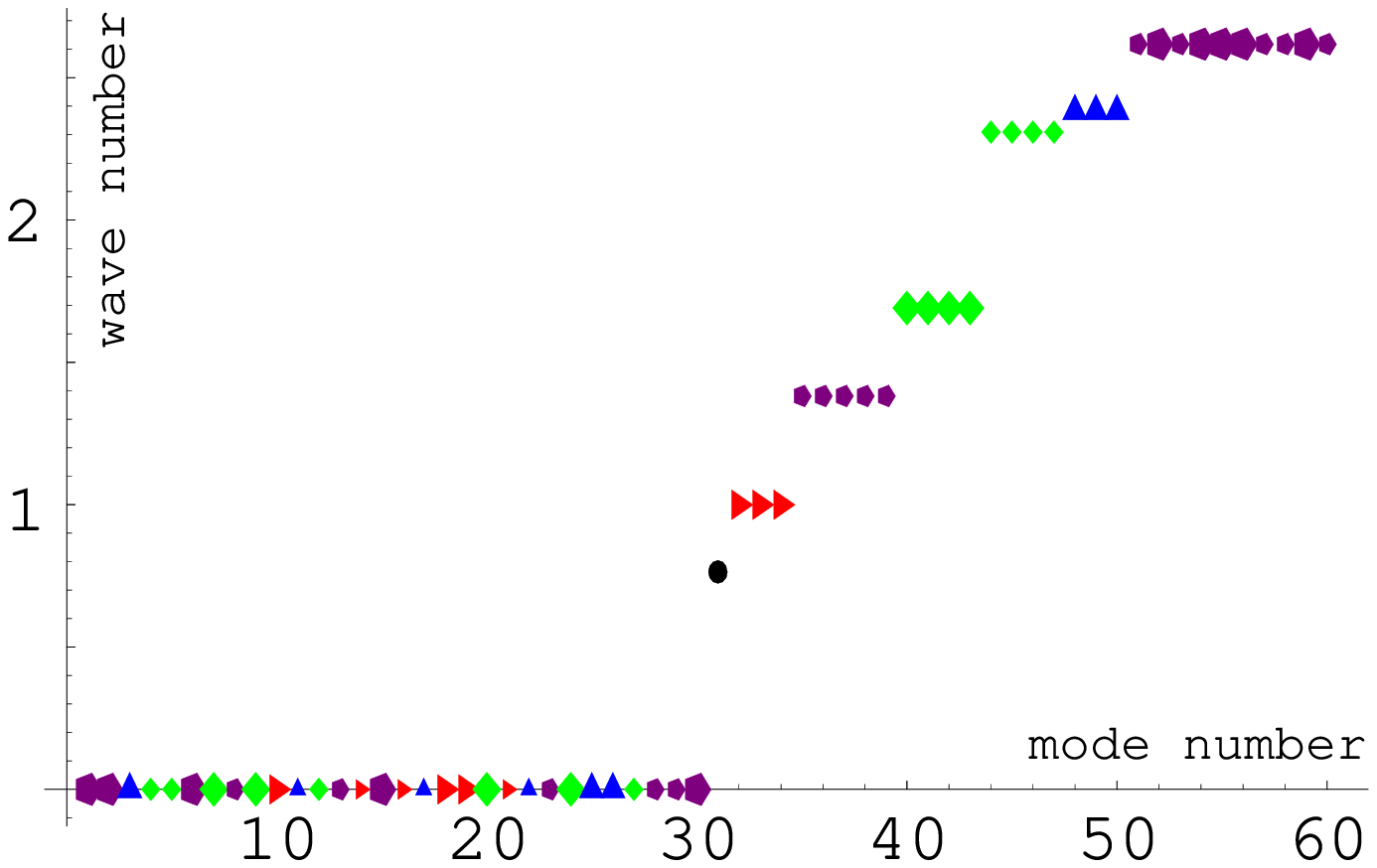}\\
\end{center}
\caption{\em{Frequencies of normal modes of vibration for (a) an
    icosahedral  cage, (b) a dodecahedral cage. The modes $\rhd$
    resp.~$\triangle$ 
    belong to 3-dimensional irreducible representations
    $\Gamma^3_\pm$ resp.~$\Gamma^{3'}_\pm$ of the icosahedral
    group. The diamond (resp.~pentagon) modes belong to 4
    (resp.~5)-dimensional irreducible representations. Small (large)
    symbols refer to even (odd) parity. The $x$-axis
    labels the normal modes while the $y$-axis gives the wave numbers
    up to an overall normalisation.}}
\label{fig:spectra}\end{figure}

Standard group theoretical methods reviewed in~\cite{PT:2008a}
are used to calculate the decomposition of the 36-dimensional
displacement representation $\Gamma^{\text{displ},36}_{\text{ICO}}$ of
the vertices into irreducible representations of the full icosahedral
group ${\cal I}_h$, which contains 60 proper rotations and an extra 60
elements obtained by multiplication of the latter by the inversion
operation. The result is (see also~\cite{Widom:2007a}),
\begin{equation} \label{ico}
\Gamma^{\text{displ},36}_{\text{ICO}}= \Gamma^1_+ + \Gamma^3_++2\Gamma^3_- +\Gamma^{3'}_-+\Gamma^4_+ +\Gamma^4_-+2\Gamma^5_+ +\Gamma^5_-\, ,
\end{equation}
where the numerical superscripts indicate the dimensionality of the
corresponding irreducible representation, while the $\pm$ subscripts
refer to different irreducible representations of same
dimension. If the subgroup ${\cal I}$ of 60 proper rotations
  is used instead, $\pm$ representations are indistinguishable. The
same group theoretical method yields, for the motion of the
dodecahedral cage, the 60-dimensional displacement representation
$\Gamma^{\text{displ},60}_{\text{DODE}}$,
\begin{multline}\label{dode}
\Gamma^{\text{displ},60}_{\text{DODE}}= \Gamma^1_+ + \Gamma^3_++2\Gamma^3_- +\Gamma^{3'}_-+\Gamma^4_+ +\Gamma^4_-+2\Gamma^5_+ +\Gamma^5_-
\\ +\Gamma^{3'}_+ +\Gamma^{3'}_- +\Gamma^4_+ +\Gamma^4_- +\Gamma^5_+ +\Gamma^5_-\, .
\end{multline}
Although Eq.(\ref{ico}) depends only on icosahedral symmetry and not
on actual links between icosahedral vertices, it is useful to
visualise an icosahedral cage formed by point masses at its vertices
joined by identical springs along its 30 edges. In contradistinction
with the dodecahedral cage, the icosahedral cage would have no
non-trivial zero-modes, in accordance with the fact that the number of
its genuine degrees of freedom $3\times 12 -6=30$ is equal to the
number of its constraints. The spectra of the icosahedral and
dodecahedral cages are depicted in Fig.\,\ref{fig:spectra}(a)$\&$(b).

\begin{figure}[ht]
\begin{center}
\vspace{-2ex}
\includegraphics[width=8.8cm,keepaspectratio]{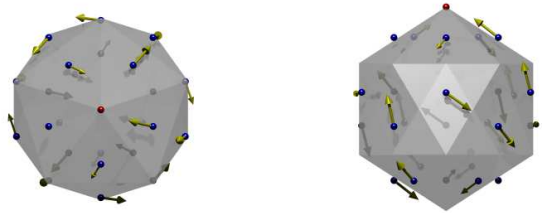}
\end{center}
\vspace{-5ex}
\caption{\em{Top and side views of the normal modes of a dodecahedral system belonging to the irreducible representation $\Gamma^{3'+}$ of the icosahedral group ${\cal I}$. }}
\label{fig:3'+}\end{figure}

As no icosahedral motion leaves the dual dodecahedron fixed, the
icosahedral motions induce a 36-dimensional vector space of
(infinitesimal) motions of the dual dodecahedron.  These include the
six global zero-modes $\Gamma^3_+ +\Gamma^3_-$ which are identical in
both systems (as well as the global dilation vibrational mode
$\Gamma^1_+$). All icosahedral normal modes of vibration pertaining to
an irreducible representation of ${\cal I}_h$ in
$\Gamma^{\text{displ},36}_{\text{ICO}}$ must be linear combinations of
dodecahedral normal modes belonging to the same irreducible
representation. Hence, provided that the vibrational modes of the
icosahedron have non-vanishing components in the vibrational modes of
finite frequency of the dodecahedron, as is true by inspection, the
set of representations in \eqref{dode} which are not contained in
\eqref{ico} describe the non-trivial zero-modes of the dodecahedron.
It follows that a capsid whose proteins would be modelled by 20~point
masses located at the centre of each face of an icosahedron, and with
30~springs organised in a network modelled at equilibrium by a
dodecahedron, develops 24 zero-modes of vibration which are organised
in the following irreducible representations of the icosahedral group,
\begin{equation} 
\label{zeromodes}
\Gamma^{3'}_++\Gamma^{3'}_- +\Gamma^4_+ +\Gamma^4_- +\Gamma^5_+ +\Gamma^5_-\, .
\end{equation}
Note that the zero-modes belonging to the $\Gamma^{3'}_+$
representation appear in \eqref{dode} but have no counterpart in
\eqref{ico}. Hence they must induce no motion when mapped back to an
icosahedral system, that is, the sum of the displacements of any 5
vertices defining a dodecahedral face must be zero.  This is indeed
the case as illustrated in Fig.\,\ref{fig:3'+}, where the icosahedral
system may be thought of as constructed from the centre-of-mass
positions of the dodecahedron vertices of each pentagon face.  The
stabilisation of such a capsid requires the introduction of at least
24 further springs in a manner that respects the icosahedral
symmetry. The magnitudes of the spring constants determine how much
the zero-modes are lifted from zero, and shape the structure of the
created low-frequency plateau.

Despite its simplicity, the approximation just described catches an
essential feature of the Caspar-Klug viruses, namely the 24-state
low-energy plateau with group content~\eqref{zeromodes}.  Indeed,
consider again a $T=n$ capsid with $3n$ proteins per icosahedral face
but now treat them, in accordance with our introductory discussions,
as $3n$ point masses located at the corresponding protein's centre of
mass calculated from the PDB files. We link them by springs according
to the association energies listed in VIPER~\footnote{In
  \cite{PT:2008a} the point-mass coordinates were projected radially
  onto icosahedron faces, which blurs stability issues.}. Neglecting
first the links between different faces, we would expect a number of
capsid zero-modes, $N_0= (9n- 3k)\cdot 20 = 60\,(3n-k)$.  Here the
$9n$ degrees of freedom within a face are constrained only by $3k$
links, $k$ integer, as a consequence of the global 3-fold symmetry of
the icosahedron whose axis passes through the centre of the
icosahedral face. A stable capsid is characterised by a force matrix
having exactly six zero-modes, so that in principle, one needs to
introduce at least $60(3n-k) -6$ independent constraints (i.e.~bonds)
to stabilise the whole capsid. How these must be chosen in a
3-dimensional context is a mathematical question which requires
further investigation. It is however interesting to note that the
VIPER website gives association energies for some inter-face proteins,
which correspond to `[icosahedron] edge-crossing' bonds. By symmetry,
these come as multiples of 30 or 60 when considering the capsid as a
whole. We thus expect that, keeping the $2(3n-k)-1$ edge-crossing
bonds per edge with the largest association energies, would yield a
capsid with 24 non-trivial zero modes. In all stable coarse-grained capsids we
studied~\cite{PT:2008b}, this indeed happens, and adding extra
edge-crossing bonds lifts the 24 zero-modes appearing in
\eqref{zeromodes} to a low-frequency plateau. Our earlier
approximation thus simply amounts to replacing $3n-k$ by unity in the
above analysis.

\vspace{-2ex}
\section{Examples}

\begin{figure}[t]
\begin{center}
\vspace{1ex}
(a) \hspace{-.3cm}\includegraphics[width=3.2cm]{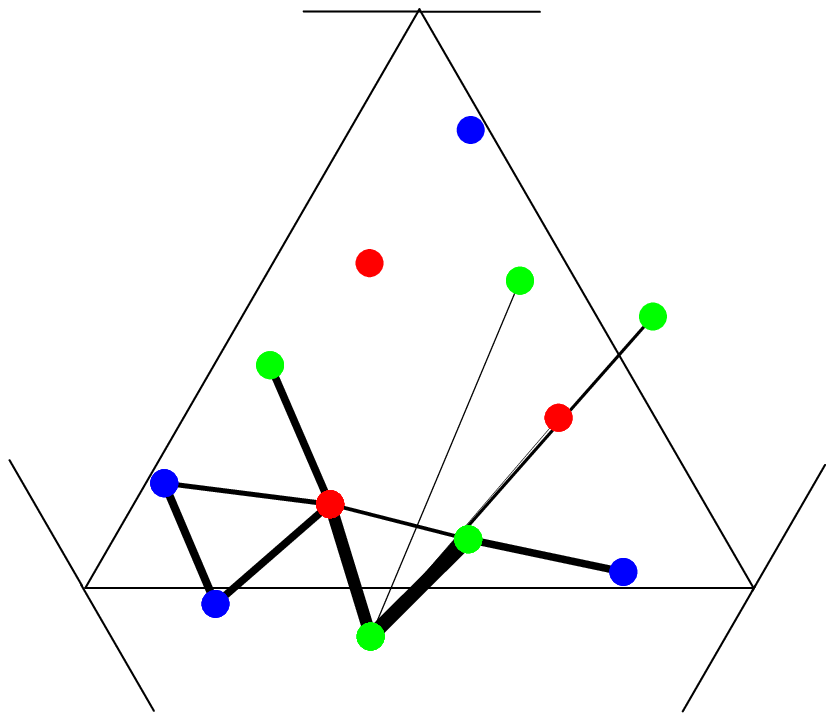}%
\,(b)  \includegraphics[width=4.3cm]{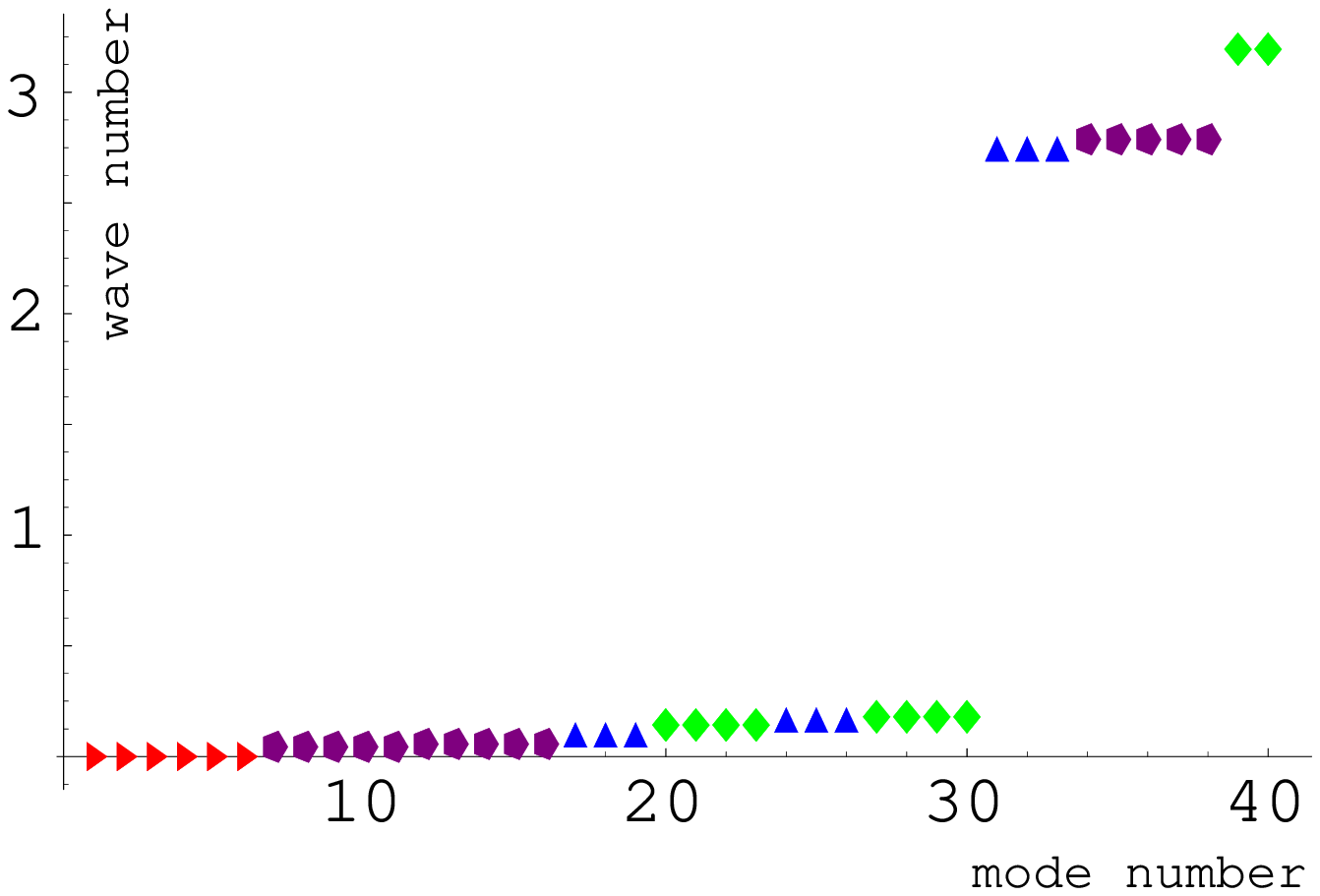}
\end{center}
\caption{\em{(a) Bond structure of RYMV as given in VIPER, (b) Its
    forty lowest frequency modes  showing a plateau of 24 near-zero
    modes. Symbols and axes as in figure~\ref{fig:spectra}.}}
\label{fig:RYMV}\end{figure}

\begin{figure}[t]
\vspace{2ex}
\begin{center}
(a) \hspace{-.3cm}\includegraphics[width=3.2 cm]{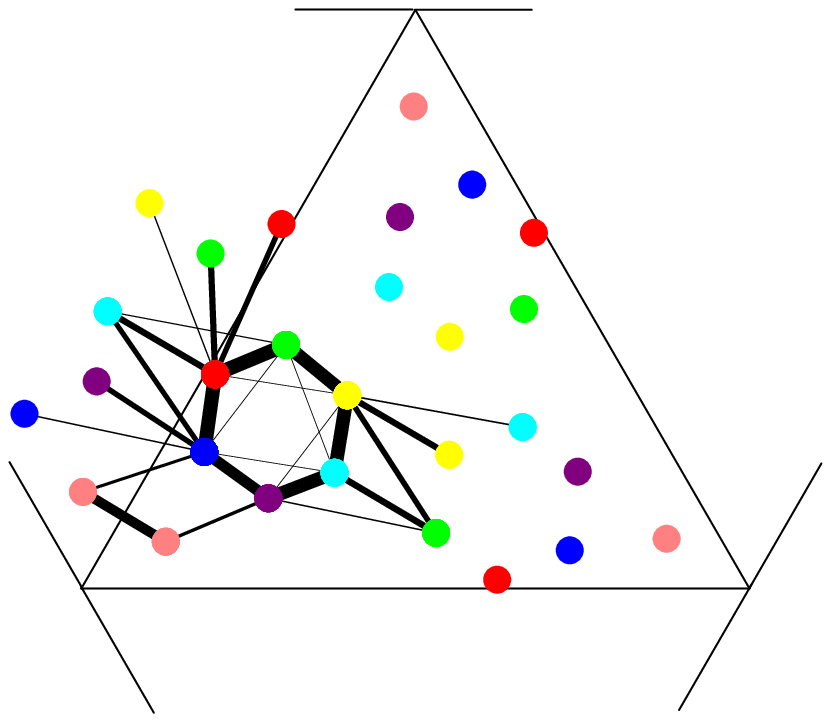}%
\,(b)  \includegraphics[width=4.3cm]{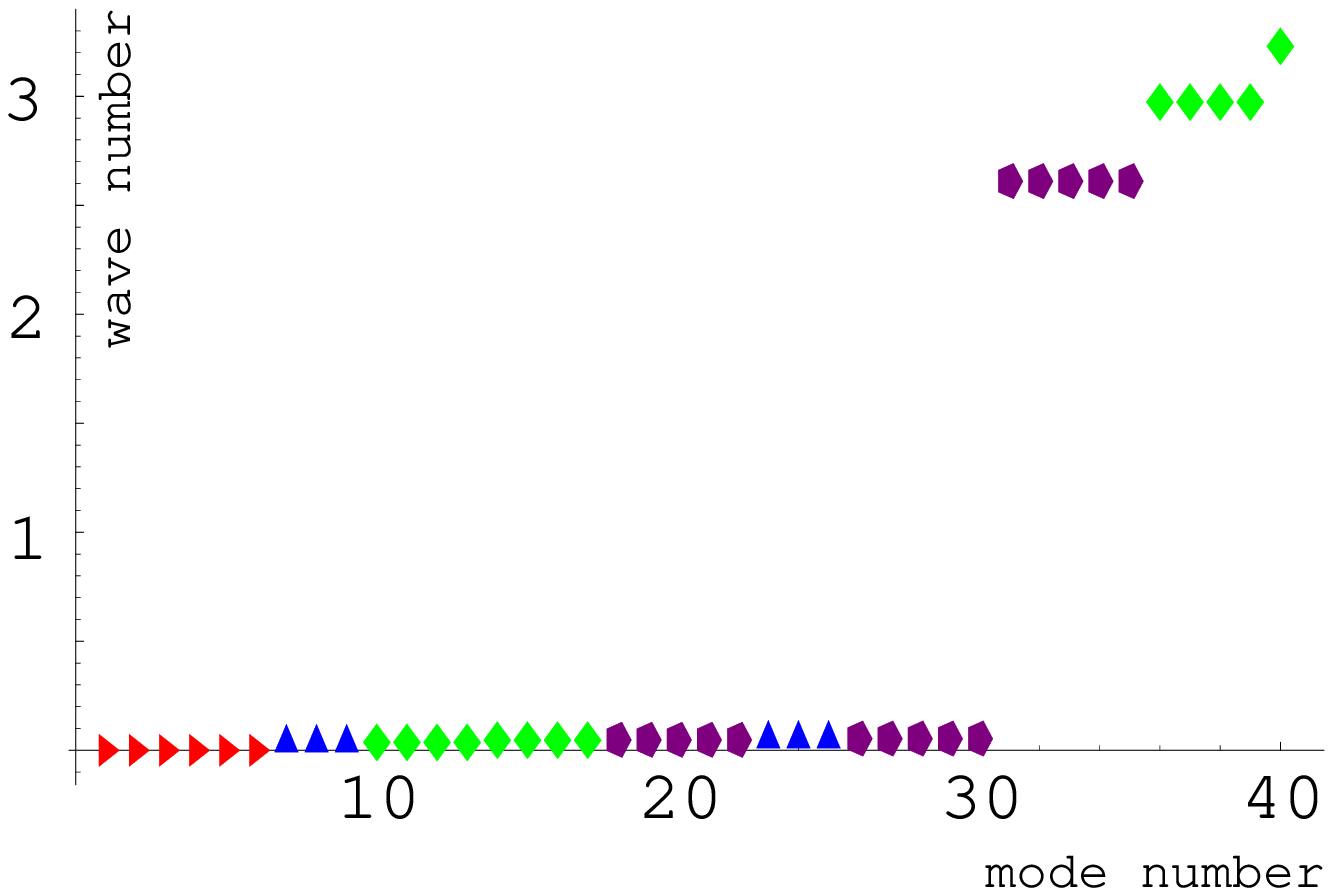}
\end{center}
\caption{\em{(a) Bond structure of HK97 as given in VIPER, (b) Its
    forty lowest frequency modes  showing a plateau of 24 near
    zero-modes with the representation content of
    \eqref{zeromodes}. Symbols and axes as in figure~\ref{fig:spectra}.}}
\label{fig:HK97}\vspace{-2ex}\end{figure}

We now illustrate our considerations with two examples. The VIPER data
for the $T=3$ Rice Yellow Mottle Virus (PDB:~1f2n) is sufficient to
stabilise the capsid. It is reproduced schematically in
Fig.\,\ref{fig:RYMV} together with the 40
lowest-frequency normal modes of vibration. There are actually 11 bonds in
  Fig.\,\ref{fig:RYMV}, but 3 of them are superimposed (2 linking
  green C-chains, and one linking a C- and a B-chain (red)). The total
  number of bonds is obtained from the figure using the global 2-, 3-
  and 5-fold symmetries of the capsid. The capsids in our
examples are not invariant under inversion, hence the symmetry group
is ${\cal I}$, and the irreducible representations are all of type
`+'. Discarding the two weakest bonds (which are both edge-crossing)
leads to~$k=4$ and $9$ edge-crossing bonds per edge. In agreement with
$2(3n-k)-1= 9$ we then observe 24 non-trivial zero modes. Restoring
the weak long-range arms which stretch from the (green) C-chains has
the crucial effect of lifting the 24 zero modes to produce a
low-frequency plateau. The $T=7\ell$ Hong-Kong 97 bacteriophage
(PDB:~2fte) has a stable capsid too. Keeping all but the six weakest
bonds (only one of them is edge-crossing in
  Fig.\,\ref{fig:HK97}) leads to $k=12$ and $17$ edge-crossing bonds
per edge, in accordance with $2(3n-k)-1= 17$, so that there are 24
non-trivial zero modes. The weak bonds serve to lift these zero modes,
and the spectrum indeed exhibits again a very low-frequency plateau of
24 modes (see Fig.\,\ref{fig:HK97}).  A variety of other capsids with
similar spectral signatures are presented in~\cite{PT:2008b}. 

\vspace{-2ex}
\section{Conclusion and discussion}

We have analysed a model of icosahedral virus capsids which
incorporates areas of varying flexibility. We have given a simple
mathematical argument which explains the appearance of a low-frequency
plateau of 24 states in the vibrational spectrum of Caspar-Klug
viruses. Our top-down approach should be viewed as a first step in
bridging the gap with the few molecular dynamics analyses published on
viral capsids, and our results should be viewed as a remarkable
property of the lowest order approximation to virus capsid vibrations.

We wish to stress that if the amplitudes of the very low frequency
modes were known, they could help decide whether they trigger
instabilities and some of the conformational changes observed
experimentally. Such changes play a fundamental role in the function
of viruses, and a fundamental explanation of their origin is still
lacking. It is moreover widely believed that the only motions that
change the capsid surface, and hence influence significantly the
interactions of the virus with the environment, are those global large
amplitude, slow motions. Hence the importance of developing models
that pin them down as accurately as possible. Another application of
their knowledge is to provide experimenters with systematics on how to
tune laser pulses in the near-infrared to produce damage on viral
capsids by forced resonance, an innovative technique developed
recently~\cite{Tsen:2007}.

We believe that our model can be developed beyond the crude
approximation we have used in our paper, and would then greatly
benefit from getting data on the absolute strength of the bonds, which
would provide us with quantitative information on low frequency
motions.  In order to test our results, however, improved techniques
are required as there are, at present, no experimental setups which
could reliably and directly check our frequencies predictions (as the
frequencies we target are too low for commonly used techniques).

\vspace{-2ex}
\section*{Acknowledgements}

The work of F.E.~was partly supported by IISN-Belgium (convention
4.4511.06 and convention 4.4505.86), and by the Belgian Federal
Science Policy Office through the Interuniversity Attraction
Pole~P~VI/11. The work of K.P.~was partly supported by VIDI grant
016.069.313 from the Dutch Organisation for Scientific Research (NWO).
Both thank the Department of Mathematical Sciences at Durham University
for its warm hospitality.


\begin{thebibliography}{29}
\expandafter\ifx\csname natexlab\endcsname\relax\def\natexlab#1{#1}\fi
\expandafter\ifx\csname bibnamefont\endcsname\relax
  \def\bibnamefont#1{#1}\fi
\expandafter\ifx\csname bibfnamefont\endcsname\relax
  \def\bibfnamefont#1{#1}\fi
\expandafter\ifx\csname citenamefont\endcsname\relax
  \def\citenamefont#1{#1}\fi
\expandafter\ifx\csname url\endcsname\relax
  \def\url#1{\texttt{#1}}\fi
\expandafter\ifx\csname urlprefix\endcsname\relax\def\urlprefix{URL }\fi
\providecommand{\bibinfo}[2]{#2}
\providecommand{\eprint}[2][]{\url{#2}}

\bibitem[{\citenamefont{Smith}(2000)}]{Smith:2000}
\bibinfo{author}{\bibfnamefont{J.~C.} \bibnamefont{Smith}},
  \emph{\bibinfo{title}{Structure and Dynamics of Biomolecules}}
  (\bibinfo{publisher}{Oxford University Press}, \bibinfo{year}{2000}), pp.
  \bibinfo{pages}{161--180}.

\bibitem[{\citenamefont{McCammon et~al.}(1976)\citenamefont{McCammon, Gelin,
  Karplus, and Wolynes}}]{Karplus:1976}
\bibinfo{author}{\bibfnamefont{J.~A.} \bibnamefont{McCammon}},
  \bibinfo{author}{\bibfnamefont{B.~R.} \bibnamefont{Gelin}},
  \bibinfo{author}{\bibfnamefont{M.}~\bibnamefont{Karplus}}, \bibnamefont{and}
  \bibinfo{author}{\bibfnamefont{P.}~\bibnamefont{Wolynes}},
  \bibinfo{journal}{Nature} \textbf{\bibinfo{volume}{262}},
  \bibinfo{pages}{325} (\bibinfo{year}{1976}).

\bibitem[{\citenamefont{Noguti and Go}(1982)}]{Noguti:1982a}
\bibinfo{author}{\bibfnamefont{T.}~\bibnamefont{Noguti}} \bibnamefont{and}
  \bibinfo{author}{\bibfnamefont{N.}~\bibnamefont{Go}},
  \bibinfo{journal}{Nature} \textbf{\bibinfo{volume}{296}},
  \bibinfo{pages}{776} (\bibinfo{year}{1982}).

\bibitem[{\citenamefont{Brooks and Karplus}(1983)}]{Brooks:1983a}
\bibinfo{author}{\bibfnamefont{B.~R.} \bibnamefont{Brooks}} \bibnamefont{and}
  \bibinfo{author}{\bibfnamefont{M.}~\bibnamefont{Karplus}},
  \bibinfo{journal}{Proc.\ Nat.\ Acad.\ Sci.} \textbf{\bibinfo{volume}{80}},
  \bibinfo{pages}{6571} (\bibinfo{year}{1983}).

\bibitem[{\citenamefont{Go et~al.}(1983)\citenamefont{Go, Noguti, and
  Nishikawa}}]{Go:1983}
\bibinfo{author}{\bibfnamefont{N.}~\bibnamefont{Go}},
  \bibinfo{author}{\bibfnamefont{T.}~\bibnamefont{Noguti}}, \bibnamefont{and}
  \bibinfo{author}{\bibfnamefont{T.}~\bibnamefont{Nishikawa}},
  \bibinfo{journal}{Proc.\ Nat.\ Acad.\ Sci.} \textbf{\bibinfo{volume}{80}},
  \bibinfo{pages}{3696} (\bibinfo{year}{1983}).

\bibitem[{\citenamefont{Levitt et~al.}(1983)\citenamefont{Levitt, Sander, and
  Stern}}]{Levitt:1983}
\bibinfo{author}{\bibfnamefont{M.}~\bibnamefont{Levitt}},
  \bibinfo{author}{\bibfnamefont{C.}~\bibnamefont{Sander}}, \bibnamefont{and}
  \bibinfo{author}{\bibfnamefont{P.}~\bibnamefont{Stern}},
  \bibinfo{journal}{Int.\ J.\ Quant.\ Chem.} \textbf{\bibinfo{volume}{10}},
  \bibinfo{pages}{181} (\bibinfo{year}{1983}).

\bibitem[{\citenamefont{Harrison}(1984)}]{Harrison:1984}
\bibinfo{author}{\bibfnamefont{R.}~\bibnamefont{Harrison}},
  \bibinfo{journal}{Biopolymers} \textbf{\bibinfo{volume}{23}},
  \bibinfo{pages}{2943} (\bibinfo{year}{1984}).

\bibitem[{\citenamefont{Elber and Karplus}(1987)}]{Karplus:1987}
\bibinfo{author}{\bibfnamefont{R.}~\bibnamefont{Elber}} \bibnamefont{and}
  \bibinfo{author}{\bibfnamefont{M.}~\bibnamefont{Karplus}},
  \bibinfo{journal}{Science} \textbf{\bibinfo{volume}{235}},
  \bibinfo{pages}{318} (\bibinfo{year}{1987}).

\bibitem[{\citenamefont{Frauenfelder et~al.}(1988)\citenamefont{Frauenfelder,
  Parak, and Young}}]{Frauenfelder:1988}
\bibinfo{author}{\bibfnamefont{H.}~\bibnamefont{Frauenfelder}},
  \bibinfo{author}{\bibfnamefont{F.}~\bibnamefont{Parak}}, \bibnamefont{and}
  \bibinfo{author}{\bibfnamefont{R.~D.} \bibnamefont{Young}},
  \bibinfo{journal}{Annu.\ Rev.\ Biophys.\ Biophys.\ Chem.}
  \textbf{\bibinfo{volume}{17}}, \bibinfo{pages}{451} (\bibinfo{year}{1988}).

\bibitem[{\citenamefont{Hong et~al.}(1990)}]{Hong:1990}
\bibinfo{author}{\bibfnamefont{M.~K.} \bibnamefont{Hong}} \bibnamefont{et~al.},
  \bibinfo{journal}{Biophys.\ J.} \textbf{\bibinfo{volume}{58}},
  \bibinfo{pages}{429} (\bibinfo{year}{1990}).

\bibitem[{\citenamefont{Brooks and Karplus}(1985)}]{Brooks:1985}
\bibinfo{author}{\bibfnamefont{B.~R.} \bibnamefont{Brooks}} \bibnamefont{and}
  \bibinfo{author}{\bibfnamefont{M.}~\bibnamefont{Karplus}},
  \bibinfo{journal}{Proc.\ Nat.\ Acad.\ Sci.} \textbf{\bibinfo{volume}{82}},
  \bibinfo{pages}{4995} (\bibinfo{year}{1985}).

\bibitem[{\citenamefont{Gibrat and Go}(1990)}]{Gibrat:1990}
\bibinfo{author}{\bibfnamefont{J.~F.} \bibnamefont{Gibrat}} \bibnamefont{and}
  \bibinfo{author}{\bibfnamefont{N.}~\bibnamefont{Go}},
  \bibinfo{journal}{Proteins} \textbf{\bibinfo{volume}{8}},
  \bibinfo{pages}{258} (\bibinfo{year}{1990}).

\bibitem[{\citenamefont{Marques and Sanejouand}(1995)}]{Marques:1995}
\bibinfo{author}{\bibfnamefont{O.}~\bibnamefont{Marques}} \bibnamefont{and}
  \bibinfo{author}{\bibfnamefont{Y.~H.} \bibnamefont{Sanejouand}},
  \bibinfo{journal}{Proteins} \textbf{\bibinfo{volume}{23}},
  \bibinfo{pages}{557} (\bibinfo{year}{1995}).

\bibitem[{\citenamefont{Mouawad and Perahia}(1996)}]{Mouawad:1996}
\bibinfo{author}{\bibfnamefont{L.}~\bibnamefont{Mouawad}} \bibnamefont{and}
  \bibinfo{author}{\bibfnamefont{D.}~\bibnamefont{Perahia}},
  \bibinfo{journal}{J.\ Mol.\ Biol.} \textbf{\bibinfo{volume}{258}},
  \bibinfo{pages}{393} (\bibinfo{year}{1996}).

\bibitem[{\citenamefont{Alexandrov et~al.}(2005)\citenamefont{Alexandrov,
  Lehnert, Echols, Milburn, Engelman, and Gerstein}}]{Alexandrov:2005}
\bibinfo{author}{\bibfnamefont{V.}~\bibnamefont{Alexandrov}},
  \bibinfo{author}{\bibfnamefont{U.}~\bibnamefont{Lehnert}},
  \bibinfo{author}{\bibfnamefont{N.}~\bibnamefont{Echols}},
  \bibinfo{author}{\bibfnamefont{D.}~\bibnamefont{Milburn}},
  \bibinfo{author}{\bibfnamefont{D.}~\bibnamefont{Engelman}}, \bibnamefont{and}
  \bibinfo{author}{\bibfnamefont{M.}~\bibnamefont{Gerstein}},
  \bibinfo{journal}{Protein Sci.} \textbf{\bibinfo{volume}{14}},
  \bibinfo{pages}{633} (\bibinfo{year}{2005}).

\bibitem[{\citenamefont{Simonson and Perahia}(1992)}]{Simonson:1992}
\bibinfo{author}{\bibfnamefont{T.}~\bibnamefont{Simonson}} \bibnamefont{and}
  \bibinfo{author}{\bibfnamefont{D.}~\bibnamefont{Perahia}},
  \bibinfo{journal}{Biophys.\ J.} \textbf{\bibinfo{volume}{61}},
  \bibinfo{pages}{427} (\bibinfo{year}{1992}).

\bibitem[{\citenamefont{Tama and Brooks}(2002)}]{Tama:2002a}
\bibinfo{author}{\bibfnamefont{F.}~\bibnamefont{Tama}} \bibnamefont{and}
  \bibinfo{author}{\bibfnamefont{C.~L.} \bibnamefont{Brooks}},
  \bibinfo{journal}{J.\ Mol.\ Biol.} \textbf{\bibinfo{volume}{318}},
  \bibinfo{pages}{733} (\bibinfo{year}{2002}).

\bibitem[{\citenamefont{Vlijmen and Karplus}(2005)}]{Vlijmen:2005a}
\bibinfo{author}{\bibfnamefont{H.~W. T.~v.} \bibnamefont{Vlijmen}}
  \bibnamefont{and} \bibinfo{author}{\bibfnamefont{M.}~\bibnamefont{Karplus}},
  \bibinfo{journal}{J.\ Mol.\ Biol.} \textbf{\bibinfo{volume}{350}},
  \bibinfo{pages}{528} (\bibinfo{year}{2005}).

\bibitem[{\citenamefont{Rader et~al.}(2005)\citenamefont{Rader, Vlad, and
  Bahar}}]{Rader:2005a}
\bibinfo{author}{\bibfnamefont{A.}~\bibnamefont{Rader}},
  \bibinfo{author}{\bibfnamefont{D.}~\bibnamefont{Vlad}}, \bibnamefont{and}
  \bibinfo{author}{\bibfnamefont{I.}~\bibnamefont{Bahar}},
  \bibinfo{journal}{Structure} \textbf{\bibinfo{volume}{13}},
  \bibinfo{pages}{413} (\bibinfo{year}{2005}).

\bibitem[{\citenamefont{Tirion}(1996)}]{Tirion:1996a}
\bibinfo{author}{\bibfnamefont{M.~M.} \bibnamefont{Tirion}},
  \bibinfo{journal}{Phys.\ Rev.\ Lett.} \textbf{\bibinfo{volume}{77}},
  \bibinfo{pages}{1905} (\bibinfo{year}{1996}).

\bibitem[{\citenamefont{Jeong et~al.}(2006{\natexlab{a}})\citenamefont{Jeong,
  Jang, and Kim}}]{Kim:2006a}
\bibinfo{author}{\bibfnamefont{J.~I.} \bibnamefont{Jeong}},
  \bibinfo{author}{\bibfnamefont{Y.}~\bibnamefont{Jang}}, \bibnamefont{and}
  \bibinfo{author}{\bibfnamefont{M.~K.} \bibnamefont{Kim}},
  \bibinfo{journal}{J.\ Mol.\ Graphics and Modelling}
  \textbf{\bibinfo{volume}{24}}, \bibinfo{pages}{296}
  (\bibinfo{year}{2006}{\natexlab{a}});

  \bibinfo{journal}{Int.\ J.\ Control, Automation and Systems}
  \textbf{\bibinfo{volume}{4}}, \bibinfo{pages}{382}
  (\bibinfo{year}{2006}{\natexlab{b}});

  \bibinfo{journal}{Nucl.\ Acids Res.} \textbf{\bibinfo{volume}{34}},
  \bibinfo{pages}{W57} (\bibinfo{year}{2006}{\natexlab{c}}).

\bibitem[{\citenamefont{Caspar and Klug}(1962)}]{Caspar:1962a}
\bibinfo{author}{\bibfnamefont{D.}~\bibnamefont{Caspar}} \bibnamefont{and}
  \bibinfo{author}{\bibfnamefont{A.}~\bibnamefont{Klug}},
  \bibinfo{journal}{Cold Spring Harbor Sympos.~Quant.~Biol.}
  \textbf{\bibinfo{volume}{27}}, \bibinfo{pages}{1} (\bibinfo{year}{1962}).

\bibitem[{\citenamefont{Shepherd et~al.}(2006)\citenamefont{Shepherd, Borelli,
  Lander, Natarajan, Siddavanahalli, Bajaj, Johnson, Brooks, and
  Reddy}}]{Shepherd:2006a}
\bibinfo{author}{\bibfnamefont{C.}~\bibnamefont{Shepherd}},
  \bibinfo{author}{\bibfnamefont{I.}~\bibnamefont{Borelli}},
  \bibinfo{author}{\bibfnamefont{G.}~\bibnamefont{Lander}},
  \bibinfo{author}{\bibfnamefont{P.}~\bibnamefont{Natarajan}},
  \bibinfo{author}{\bibfnamefont{V.}~\bibnamefont{Siddavanahalli}},
  \bibinfo{author}{\bibfnamefont{C.}~\bibnamefont{Bajaj}},
  \bibinfo{author}{\bibfnamefont{J.}~\bibnamefont{Johnson}},
  \bibinfo{author}{\bibfnamefont{C.~I.} \bibnamefont{Brooks}},
  \bibnamefont{and} \bibinfo{author}{\bibfnamefont{V.}~\bibnamefont{Reddy}},
  \bibinfo{journal}{Nucl.\ Acids Res.} \textbf{\bibinfo{volume}{34}},
  \bibinfo{pages}{D386} (\bibinfo{year}{2006}).

\bibitem[{\citenamefont{Peeters and Taormina}(2008{\natexlab{a}})}]{PT:2008a}
\bibinfo{author}{\bibfnamefont{K.}~\bibnamefont{Peeters}} \bibnamefont{and}
  \bibinfo{author}{\bibfnamefont{A.}~\bibnamefont{Taormina}}, 
  \bibinfo{journal}{Comp.\ Math.\ Methods in Medicine}
  \textbf{\bibinfo{volume}{9}}, \bibinfo{pages}{211}
  (\bibinfo{year}{2008}{\natexlab{a}}), 
  \eprint{arXiv:0802.2620}.

\bibitem[{\citenamefont{Widom et~al.}(2007)\citenamefont{Widom, Lidmar, and
  Nelson}}]{Widom:2007a}
\bibinfo{author}{\bibfnamefont{M.}~\bibnamefont{Widom}},
  \bibinfo{author}{\bibfnamefont{J.}~\bibnamefont{Lidmar}}, \bibnamefont{and}
  \bibinfo{author}{\bibfnamefont{D.~R.} \bibnamefont{Nelson}},
  \bibinfo{journal}{Phys.\ Rev.} \textbf{\bibinfo{volume}{E76}},
  \bibinfo{pages}{031911} (\bibinfo{year}{2007}), \eprint{arXiv:0706.4291}.

\bibitem[{\citenamefont{Peeters and Taormina}(2008{\natexlab{b}})}]{PT:2008b}
\bibinfo{author}{\bibfnamefont{K.}~\bibnamefont{Peeters}} \bibnamefont{and}
  \bibinfo{author}{\bibfnamefont{A.}~\bibnamefont{Taormina}}
  \emph{\bibinfo{title}{Group theory of icosahedral virus capsids: a dynamical top-down approach}}
  (\bibinfo{year}{2008}{\natexlab{b}}),  \eprint{arXiv:0806.1029}.

\bibitem[{\citenamefont{Tsen et~al.}(2007)\citenamefont{Tsen, Tsen, Sankey, and
  Kiang}}]{Tsen:2007}
\bibinfo{author}{\bibfnamefont{K.~T.} \bibnamefont{Tsen}},
  \bibinfo{author}{\bibfnamefont{S.-W.~D.} \bibnamefont{Tsen}},
  \bibinfo{author}{\bibfnamefont{O.~F.} \bibnamefont{Sankey}},
  \bibnamefont{and} \bibinfo{author}{\bibfnamefont{J.~G.} \bibnamefont{Kiang}},
  \bibinfo{journal}{J.\ Phys.: Cond.\ Mat.} \textbf{\bibinfo{volume}{19}},
  \bibinfo{pages}{472201} (\bibinfo{year}{2007}).

\end{thebibliography}

\end{document}